# BAYESIAN UPDATING OF SEISMIC GROUND FAILURE ESTIMATES VIA CAUSAL GRAPHICAL MODELS AND SATELLITE IMAGERY

S. Xu[1], J. Dimasaka[2], D. J. Wald[3], H. Noh[4]

[1] *Assistant Professor, State University of New York at Stony Brook, susu.xu@stonybrook.edu*
[2] *Graduate Student, Stanford University, dimasaka@stanford.edu*
[3] *Seismologist, U.S. Geological Survey, wald@usgs.gov*
[4] *Associate Professor, Stanford University, noh@stanford.edu*

## *Abstract*

Earthquake-induced secondary ground failure hazards, such as liquefaction and landslides, result in catastrophic building and infrastructure damage as well as human fatalities. To facilitate emergency responses and mitigate losses, the U.S. Geological Survey provides a rapid hazard estimation system for earthquake-triggered landslides and liquefaction using geospatial susceptibility proxies and ShakeMap ground motion estimates. However, the resolution and accuracy of these models are often limited by coarse granularity and large uncertainties of available geospatial features provided at a regional scale. Recently, with the advancement of remote sensing technologies, synthetic aperture radar (SAR) images are captured and analyzed to obtain a rapid estimate of earthquake-induced correlation changes between pre- and post-event images. These correlation changes indicate ground failures and building damage, showing the potential to provide supplementary information for rapid hazard and loss estimation. However, the exact causes of changes in satellite images are not directly ascertained by the damage proxy maps (DPMs) alone. For example, changes could be due to building damage, landslides, liquefaction, noise, or any combination thereof. More importantly, the occurrence and intensity of landslides, liquefaction, and building damages are spatially correlated, which makes it more challenging to distinguish the sources of any such changes.

In this study, we develop a generalized causal graph-based Bayesian network that models the physical interdependencies between geospatial features, seismic ground failures, and building damage, as well as DPMs. Geospatial features provide physical insights for estimating ground failure occurrence while DPMs contain event-specific surface change observations. This physics-informed causal graph incorporates these variables with complex physical relationships in one holistic Bayesian updating scheme to effectively fuse information from both geospatial models and remote sensing data. This framework is scalable and flexible enough to deal with highly complex multi-hazard combinations. We then develop a stochastic variational inference algorithm to jointly update the intractable posterior probabilities of unobserved landslides, liquefaction, and building damage at different locations efficiently. In addition, a local graphical model pruning algorithm is presented to reduce the computational cost of large-scale seismic ground failure estimation. We apply this framework to the September 2018 Hokkaido Iburi-Tobu, Japan (M6.6) earthquake and January 2020 Southwest Puerto Rico (M6.4) earthquake to evaluate the performance of our algorithm.

## 1. Introduction

Earthquake-induced ground failures and building damage cause significant economic losses and fatalities. For example, the 2008 Wenchuan earthquake in China triggered about 200,000 landslides, leading to around 26,500 deaths [1] in addition to nearly 60,000 shaking-induced fatalities. The series of earthquakes in Christchurch, New Zealand, in 2011, induced liquefaction over one third of the city, affecting more than 6,000 buildings and resulting in huge economic costs [2]. These hazards have been shown to cause disruption to lifelines and structural damage to buildings [3]. Therefore, rapidly and accurately localizing and estimating ground failure and damage occurrences are beneficial to effective and efficient response and recovery.

Various approaches have been developed over the years for estimating the location and intensity of earthquake-induced ground failures and building damage. Traditional methods include physical [4, 5, 6] and statistical models [7, 8, 9, 10]. Physical models such as the Newmark displacement-based landslide model [5, 6] and liquefaction potential index [11] are often based on fundamental physical processes but cannot be employed when geotechnical data are absent. Although these models are based on the underlying physical processes, they are also often error prone due to the simplification of complex physical processes. Alternatively, statistical models can be calibrated against patterns of past ground failures using historical inventories given geospatial susceptibility proxies (e.g., slope, lithology.) and approximate ground motion levels that triggered them to estimated failure [8, 9]. The resolution and accuracy of these statistical models are often constrained by the limited availability of geospatial features as well as modeling uncertainties. For example, it is difficult to acquire comprehensive high-resolution lithology, land cover type, and other predictor variables for landslide model susceptibility, or soil strength and water depth needed for liquefaction modeling. Moreover, how these geospatial proxies impact the probability of landslide and liquefaction, together with ground shaking and moisture conditions, are spatially complex processes with large uncertainties. Significant challenges persist regarding the adaptation and generalization of statistical models trained on past inventories to new events because ground failures are sensitive to subtle environmental factors that vary from event to event.

Recently, remote sensing methods have been developed to enable a quick estimation of ground failure and damage [12, 13, 14]. Most notably, damage proxy maps (DPMs), developed by researchers of NASA Advanced Rapid Imaging and Analysis (ARIA) team, model multi-temporal correlation changes between satellite images captured before and after an earthquake to indicate earthquake-induced ground surface changes [15]. Nevertheless, it is very difficult to categorize different types of changes through these imagery data, such as ground failures and building damages, as well as noise from vegetation growth and anthropogenic modifications, especially when these changes co-occur [16]. For example, ground shaking can cause liquefaction and building damage, but it is difficult to differentiate building damage and liquefaction directly from satellite images due to spatial overlap, which significantly restricts attribution to specific phenomena.

To address the aforementioned challenges, we incorporate the geospatial models and remote sensing observations to jointly improve ground failure estimation performance. Remote sensing data provide rich information about event-specific surface change patterns, while geospatial models present *physical insights* to help distinguish different types of ground failure, damage, and noise. Some prior studies incorporated geospatial features and remote sensing observations for single-type ground failure estimation using linear combinations or black-box supervised classifiers [17, 18, 19]. However, these approaches lack consideration of complex and event-varying interdependencies between different types of ground failure, damage, remotely sensed observations, and noise, which limit their applicability for common multi-hazard, mixed-signal scenarios.

In this work, we develop a new Bayesian updating framework integrating geospatial models with remote sensing observations through a physics-informed causal graph to achieve fast and effective joint estimation of regional ground failure and building damage. This framework is based on a Bayesian network that uses graph-based representation as the foundation for encoding a set of conditional dependency relationships, e.g., physical causal relationship, among multiple random variables. The random variables are represented as nodes, while their causal relationships are modeled through edges with directions. The posterior distributions of these random variables can be obtained by learning their conditional dependencies through Bayesian updating. The Bayesian network is shown to be a powerful tool for deciphering complex causation among multiple variables from a group of data [20, 21, 22]. In the context of this work, we build a causal graph that models physical interdependencies among landslides, liquefaction, building damage, and DPMs, which allows us to construct more complex and nonlinear relationships among different variables and thus more accurately approximate their physical relationships. Explicitly modeling the interdependencies via the causal graph enables a more comprehensive physical reasoning of the image changes in DPMs as ground failures, building damages, environmental noises, or all of them, and thus reduces bias and uncertainties in estimations.

Specifically, we construct a Bayesian network based on the causal graph and develop a stochastic variational inference algorithm to infer location-specific posterior distributions of ground failure and building damage. The main challenges for the posterior distribution estimation on this Bayesian network are that (1) in near-real time, all ground failures and building damage are typically unobserved without any ground truth information; and (2) the statistical relationship between ground failure, building damage, and remote sensing observations are complex and unknown. These challenges make it difficult to infer the posterior distributions of unobserved variables. To address these challenges, we develop a stochastic variational inference algorithm to approximate the posterior distributions of unobserved ground failure and building damage and their statistical correlations by maximizing the lower bound of the likelihood of observed DPMs. The first step of the stochastic variational inference involves randomly sampling small batches of locations from the entire macroseismic zone to enable scalability of the method. We then jointly approximate the posterior of unobserved variables related to sampled locations with variational inference. Variational inference aims to derive a lower bound for the marginal likelihood of observed variables to provide analytical approximation of posterior distributions of unobserved variables with complex statistical correlations [23, 24, 25]. We then conduct a stochastic gradient descent to update the parameters representing statistical correlations between different variables. As the algorithm converges, we uncover the optimal combination of ground failure and building damage posteriors as well as their statistical relationships in an efficient way. With location-specific image-change information from remote sensing observations, we estimate ground failures and building damage with higher resolution and accuracy.

There are three main contributions of this work: (1) We introduce a new physics-informed causal graph-based model to integrate empirical statistical models, physical interdependencies, and remote sensing observations for large-scale joint ground failure and building damage estimation. This model provides a physical interpretation of DPM changes and higher spatial resolution and accuracy than the prior model, as well as a new approach to fuse different data modalities for seismic hazard estimation. (2) We design a stochastic variational inference framework to jointly approximate posterior probability of ground failures and building damage from complex Bayesian networks, without the need for any ground truth labels. We develop variational bounds that are applicable to a family of hierarchical Bayesian networks composed of unobserved binary variables and observed exponential family variables. We also enable scalability and computational efficiency for large-scale graph inference through stochastic gradient descent. (3) We evaluate our method on earthquakes in Hokkaido, Japan, and Puerto Rico, USA, for which some ground failure and building damage observations are available for validation to show the effectiveness of the framework in different types of events and environments.

The remaining sections of this paper are organized as follows: In Section 2, we introduce our method, including the causal graph-based Bayesian network, stochastic variational inference, as well as the final algorithm of Bayesian updating. In Section 3, we evaluate our algorithm using the aforementioned case studies and discuss its performance.

## 2. Causal Graph-based Bayesian Network for Updating Ground Failure Estimation

In this section, we introduce our causal graph-based Bayesian network to integrate ground failures, building damage, remote sensing observations, and their interdependencies. We then present a stochastic variational inference approach to approximate the intractable (i.e., cannot be directly modeled) posteriors of unobserved ground failure and building damage using remote sensing observations and prior geospatial models. Finally, we describe the optimization framework to find optimal weights to represent the statistical relationships between different predictors, ground failure, building damage, and environmental and anthropogenic noise, as well as remote sensing observations. In the following sections, we use DPMs mentioned above to represent our remote sensing observations.

## 2.1 Causal Graph-based Bayesian Network

In this section, we first model a causal graph-based Bayesian network in Fig. 1 to represent the statistical relationships between ground failure, building damage, and DPMs. Prior conditional distributions of nodes in the graphical models are further specified.

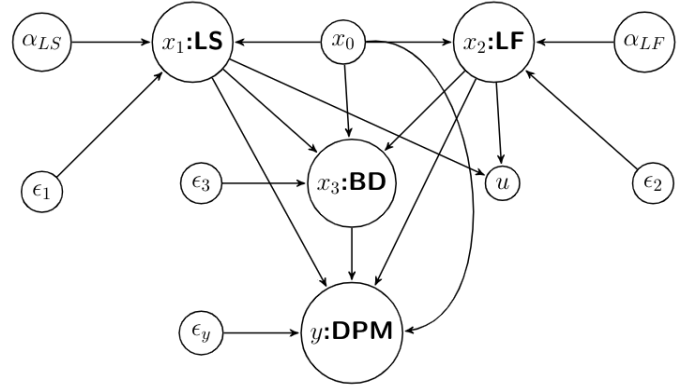

Fig. 1 – Damage causal graph depicting physical interdependencies between ground failures (landslide (LS) and liquefaction (LF)), building damages (BD), remote sensing observations (DPM), bias terms, and noises.

Given one location, we first denote $x_1$ for landslide (LS), $x_2$ for liquefaction (LF), and $x_3$ for building damage (BD). All $x$'s are unobserved binary variables and thus are assumed to have a Bernoulli distribution. The unobserved ground failure and damage nodes have binary variables $x_i \in \{0,1\}$, where $i \in \{1, 2, 3\}$. When the LS/LF/BD happen, $x_i = 1$, otherwise $x_i = 0$. We use $y$ to refer to *DPM* observation, which is a continuous variable bounded by [0,1]. We denote $\alpha_{LS}$ and $\alpha_{LF}$ as prior probabilities of LS and LF generated from the statistical models used in the U.S. Geological Survey (USGS) near-real-time ground failure product, respectively [10, 26]. We denote $\epsilon_1, \epsilon_2, \epsilon_3, \epsilon_y$ as unobserved Gaussian noise in LS, LF, and BD, as well as in DPMs, respectively. We also set up a XOR node $u$ to refer to the mutually exclusive states between its parents: landslide ($x_1$:LS) and liquefaction ($x_2$:LF) to constrain that $x_1 x_2 = 0$. This is based on our assumption that most mapped landslide and liquefaction occurrences are mutually exclusive spatially.

All nodes are linked by an arbitrary directed acyclic graph in Fig. 1. We further give quantitative definitions of these links, i.e., statistical relationships, between different random variables. For unobserved nodes, $x_i \in \{0,1\}$, where $i \in \{1,2,3\}$, we define $w_{\epsilon_i}$, $w_{\alpha_i}$, and $w_{ki}$ ($k$ is any parent node of $i$) to quantify the correlations. We also define $w_{0i}$ and $x_0 = 1$, which allows the node $i$ to be active even when its other parent nodes are inactive. For example, even if neither landsliding nor liquefaction is present, building damage is still possible due to the shaking. Using $\mathcal{P}(i)$ to represent the parents of node $i$ (excluding the leaf nodes $x_0$ and $\epsilon_i$), the conditional distribution of $x_i$ is modeled as:

$$\log \frac{p(x_i = 1 | x_{\mathcal{P}(i)}, \epsilon_i)}{1 - p(x_i = 1 | x_{\mathcal{P}(i)}, \epsilon_i)} = w_{\epsilon_i} \epsilon_i + w_{0i} x_0 + \sum_{k \in \mathcal{P}(i)} w_{ki} x_k. \tag{1}$$

The above logit relationship between LS/LF/BD and their parent nodes follows the assumption of the logistic regression model, which is used in the statistical models of LS and LF used by the USGS [10, 26]. If all parents are active ($x_k = 1; \forall k \in P(i)$), they activate the child node $i$ with probability of $\frac{1}{1 + exp(-w_{i\epsilon}\epsilon_i - w_{0i}x_0 - \sum_{k \in P(i)} w_{ki} x_k)}$, regardless of the states of other parents. If $x_k = 0$, parent $k$ has no influence on the state of $x_i$. $w_{\epsilon_i}\epsilon_i$ measures noise in the physical dependencies. $w_{0i}x_0$ helps estimate and cancel the bias introduced by prior models. The node activation probabilities are defined as follows:

$$p\left(x_i\middle|x_{\mathcal{P}(i)},\epsilon_i\right) = \left[p\left(x_i=1\middle|x_{\mathcal{P}(i)},\epsilon_i\right)\right]^{x_i} \times \left[1-p\left(x_i=1\middle|x_{\mathcal{P}(i)},\epsilon_i\right)\right]^{1-x_i}. \tag{2}$$

Similarly, given $y$, we use $\mathcal{P}(y)$ to define the parents of $y$. Based on the empirical probability density function of DPMs, we found $y$ subjects to a truncated lognormal distribution bounded by [0,1]. Therefore, we model the dependencies between $y$'s parents and $y$ as follows:

$$\log(y+\delta) = w_\epsilon\epsilon_y + w_{0y} + \sum_{i\in\mathcal{P}(y)}w_{iy}\,x_i, \tag{3}$$

where $\epsilon_y$ is a normal distribution representing the random noise in the DPM, $w_{0y}\le 0$ estimates regional bias, $\delta\to 0^+$, and $y|\mathcal{P}(y)$ has a truncated log-normal distribution. The XOR node $u$ poses a constraint on the mutually exclusive relationship between the parent nodes LS and LF. We denote the parents of $u$ as $\mathcal{P}(u)$; the distribution of $u|x_{P(u)}$ is thus a Kronecker delta function, i.e.:

$$p\left(u\middle|x_{\mathcal{P}(u)}\right) = \begin{cases} 1 \;\; if\; u = \prod_{k\in\mathcal{P}(u)}x_k \\ 0 \;\; if\; u \neq \prod_{k\in\mathcal{P}(u)}x_k \end{cases}. \tag{4}$$

It is difficult to optimize the posterior based on the discrete Kronecker delta function. We first transform this function into its continuous version, which is a Dirac delta function $\delta\left(u-\prod_{k\in\mathcal{P}(u)}x_k\right)$. Then, we use a Gaussian distribution to approximate this Dirac delta function, which is

$$p\left(u\middle|x_{\mathcal{P}(u)}\right) = \frac{1}{\sqrt{2\pi}\sigma}\exp\left[-\frac{\left(u-\prod_{k\in\mathcal{P}(u)}x_k\right)^2}{2\sigma^2}\right], \tag{5}$$

where $\sigma$ is a small real positive number, $\sigma^2\to 0$. As we assume mutual exclusivity between LS and LF always exists, $u$ is always 0.

With the above distribution and conditional distribution assumptions, we construct a Bayesian network based on the causal graph which effectively captures the dependencies between different ground failure types, building damage, and remote sensing observations. This network contains multiple unobserved random variables and unknown statistical relationships between random variables. The complex statistical dependencies make the posterior of unobserved random variables intractable. Therefore, we develop a stochastic variational inference framework to approximate the intractable posterior of unobserved ground failure and building damage.

## 2.2 Stochastic Variational Inference

With the Bayesian network constructed in Section 2.1, we need to further infer the posterior of ground failure and building damage. However, the problem is that (1) both statistical relationships, i.e., $w_i$ and distributions of ground failures and building damage are unknown, and (2) the target macroseismic region could be large, which makes it computationally costly to jointly update posteriors of ground failure/building damage over the entire map. To address this problem, variational inference is introduced to factorize the Bayesian network first and approximate the posterior distributions of unobserved variables through maximizing the log-likelihood of observed variables. To ensure the scalability of our method, variational inference is conducted on a small batch of randomly sampled locations in each iteration, described further in Section 2.3.

For each location $l$, we define a variational distribution $q\left(X^l\right)$, which further factorizes over hidden (unobserved) nodes:

$$q(X^l) = \prod_i q\left(x_i^l\right) = \prod_i \left(q_i^l\right)^{x_i^l}\left(1-q_i^l\right)^{1-x_i^l}. \tag{6}$$

At each geo-location, $q_i^l$ is defined to approximate the posterior probability that node $i$ at location $l$ is active. We fix $q_0=1$ so the node $x_0$ is always on. For any $q\left(X^l\right)$, the marginal log-likelihood of the observed DPM $y^l$ and $u$ can be lower bounded by Jensen's inequality as follows:

$$log\,P\,(y^l, u) \geq E\left[log\,\frac{p(y^l, u, X^l, \epsilon)}{q(X^l, \epsilon)}\right] \tag{7}$$

$$= \int q(X^l, \epsilon)\,log\,p\,(y^l, u, X^l, \epsilon)d(X^l, \epsilon) - \int q(X^l, \epsilon)\,log\,q\,(X^l, \epsilon)d(X^l, \epsilon) \tag{8}$$

$$= E_{q(X^l, \epsilon)}[log\,p\,(y^l, u, X^l, \epsilon)] - E_{q(X^l, \epsilon)}[log\,q\,(X^l, \epsilon)]. \tag{9}$$

Using the above equation, given a map containing a set of locations, $\mathcal{L}$, we further derive a tight lower bound for the log-likelihood as follows:

$$log\,P\,(y, u) \geq L(q, w), \tag{10}$$

where $L(q, w) =$

$$\sum_{l \in \mathcal{L}} \Bigg\{ -log\,y^l - log\,w_{\epsilon_y} - \frac{(log\,y^l)^2 + w_{0y}^2 + \sum_{k \in x_{\mathcal{P}(y^l)}} w_{ky}^2 q_k^l}{2w_{\epsilon_y}^2}$$

$$- \frac{\sum_{i,j \in x_{\mathcal{P}(y^l)}; i \neq j} 2w_{iy}w_{jy}q_i^l q_j^l - 2w_{0y}\,log\,y^l - 2(log\,y^l)\left(\sum_{k \in x_{\mathcal{P}(y^l)}} w_{ky}q_k^l\right)}{2w_{\epsilon_y}^2} - \frac{2w_{0y}\sum_{k \in x_{\mathcal{P}(y^l)}} w_{ky}q_k^l}{2w_{\epsilon_y}^2}$$

$$- \sum_{i \in \{LS, LF\}} q_i^l\,log\left[1 + exp\left(-w_{0i} - w_{\alpha_i}\alpha_i^l + \frac{w_{\epsilon_i}^2}{2}\right)\right] - \sum_{i \in \{LS, LF\}} (1 - q_i^l)\,log\left[1 + exp\left(w_{0i} + w_{\alpha_i}\alpha_i^l + \frac{w_{\epsilon_i}^2}{2}\right)\right]$$

$$- \sum_{i \in \{BD\}} q_i^l q_{\mathcal{P}(i)}^l\,log\left[1 + exp\left(-w_{0i} - w_{\mathcal{P}(i)i} + \frac{w_{\epsilon_i}^2}{2}\right)\right] - \sum_{i \in \{BD\}} q_i^l (1 - q_{\mathcal{P}(i)})\,log\left[1 + exp\left(-w_{0i} + \frac{w_{\epsilon_i}^2}{2}\right)\right]$$

$$- \sum_{i \in \{BD\}} (1 - q_i^l)q_{\mathcal{P}(i)}^l\,log\left[1 + exp\left(w_{0i} + w_{\mathcal{P}(i)i} + \frac{w_{\epsilon_i}^2}{2}\right)\right]$$

$$- \sum_{i \in \{BD\}} (1 - q_i^l)(1 - q_{\mathcal{P}(i)})\,log\left[1 + exp\left(w_{0i} + \frac{w_{\epsilon_i}^2}{2}\right)\right] - \frac{1}{2\sigma^2}\left[(u^l)^2 + (1 - 2u^l)\prod_{k \in \mathcal{P}(u^l)} q_k^l\right]$$

$$- \sum_{i \in \{LS, LF, BD\}} q_i^l\,log\,q_i^l - \sum_{i \in \{LS, LF, BD\}} (1 - q_i^l)\,log(1 - q_i^l) \Bigg\}. \tag{11}$$

The overall idea of getting the lower bound is to introduce auxiliary parameters to find the tight lower bound of each expectation. Jensen's inequality, Taylor's expansion, and Bayesian theorem are utilized [22]. With the tight lower bound of log-likelihood of DPM observations, we can further maximize the lower bound to find optimal posteriors of unobserved variables, i.e., LS, LF, and BD, as explained in the following subsection.

## 2.3 Expectation-Maximization Algorithm for Posterior Optimization

With the overall variational bound derived in Section 2.2, our final objective is to maximize the bound to find optimal combinations of posteriors and weights. As both posteriors and weights are unknown, we develop an expectation-maximization approach to achieve this. In the step of expectation, we derive closed-form update equations for local posteriors of LS, LF, and BD, i.e., $q_i^l$, by maximizing the lower bound and setting the gradients of the lower bound as 0, i.e., $\frac{\partial L(q, w)}{\partial q_i^l} = 0$. This gradient is obtained via the chain rule, and the optimal posterior follows the form below:

$$q_i^l = \frac{1}{1 + exp\left(-T\left(q_{\mathcal{P}(i)}^l, q_{\mathcal{S}(i)}^l, q_{\mathcal{C}(i)}^l, y^l, \alpha_i^l, w\right)\right)}, \tag{12}$$

where $\mathcal{P}(i)$ refers to the set of parent nodes, $\mathcal{C}(i)$ refers to the set of child nodes, $\mathcal{S}(i)$ refers to the set of spouse nodes that share the same child nodes with $i$. T is a nonlinear function that equals $\frac{\partial[L(q,w) + q_i^l\,log\,q_i^l + (1 - q_i^l)\,log(1 - q_i^l)]}{\partial q_i^l}$, determined by the prior of $i$, weights of edges associated with node $i$, and posteriors of the parent, child, and spouse nodes of $i$.

In the maximization step, we conducted stochastic gradient descent updates to estimate the optimal weights using a mini-batch of data randomly sampled from different locations. The edge weights $w^{(t)}$ at the iteration t are therefore updated as follows:

$$w^{(t+1)} = w^{(t)} + \rho A \nabla \mathcal{L}^{(t)}(w), \tag{13}$$

where $\rho$ controls the learning rate and $A$ is a preconditioner and here is set up as the identity matrix to accelerate convergence to high-likelihood models. In each iteration, we first randomly sample a mini-batch of locations from the given map. Then, the expectation step and maximization step are implemented to update the posterior estimations as well as the global weight parameters. As the model converges, the optimal posteriors of landslide, liquefaction, and building damage at each location are estimated.

## 3. Evaluation

We evaluate the causal graph-based Bayesian network with stochastic variational inference and expectation-maximization algorithm using the September 2018 Hokkaido Iburi-Tobu (M6.6) earthquake and the January 2020 Puerto Rico (M6.4) earthquake. In the following sections, we compare and discuss the spatial distribution of prior and posterior models with the available ground truth observations for these two earthquakes. The prior models refer to current earthquake-induced landslide [26] and liquefaction [10] hazards of the USGS ground failure product. The posterior models are the resulting landslide, liquefaction, and building damage models using our causal graph-based Bayesian network.

To evaluate the performance of these models, we measured true positive rate ($TPR$) and false positive rate ($FPR$) as follows:

$$\text{TPR} = \frac{TP}{TP + FN}, \text{FPR} = \frac{FP}{FP + TN}, \tag{14}$$

where $TP$, $TN$, $FN$, and $FP$ count the number of true positives, true negatives, false negatives, and false positives of both prior and posterior models, respectively. We also compared the performance of prior and posterior models using the receiver operating characteristic (ROC) curve to evaluate their ground failure identification ability. Besides binary classification accuracy, we also employed cross-entropy loss (CEL) to evaluate how similar to the true distribution the estimated distribution is. CEL is defined as follows:

$$\text{CEL} = -\frac{1}{N} \sum_{l \in N} \left[ \left( g^l \log q^l \right) + \left( 1 - g^l \right) \log \left( 1 - q^l \right) \right], \tag{15}$$

where $g$ is a binary logical variable with a value of 1 if a location has a ground truth observation and $q$ is the normalized prior or posterior estimates in the range [0,1] for all locations N. The lower the loss, the better the estimated probability distribution of landslide or liquefaction model.

### 3.1 September 2018 Hokkaido Iburi-Tobu earthquake

Two days after the landfall of Typhoon Jebi with heavy rainfall, on September 6, 2018, at 3:08 am (JST), devastating landslides and significant liquefaction were triggered in the southern region of Hokkaido, Japan, by a Mw 6.6 earthquake [27, 28]. The ARIA team generated DPMs using the SAR images from the ALOS-2 satellites of the Japan Aerospace Exploration Agency [29]. This DPM covered the towns of Atsuma and Abira, which were situated near the large-scale landslides [30]. The prior estimations of landslide and liquefaction are generated by ShakeMap in Atlas V3 [28] provided by the USGS.

Fig. 2 shows that the posterior landslide model (*Fig. 2c*), which has higher resolution (30 m), resembled the spatial distribution of ground truth observations (*Fig. 2d*) more accurately than the prior models (*Fig. 2a*) that had lower 230-m resolution [26]. The posterior model, which integrates the DPM (*Fig. 2b*) also identified more true positives with prominent peaks than the prior model. The $FPR$ decreased by 50% while the TPR increased by 379% compared to the prior model estimates. This improved performance of posterior landslide estimation minimized the CEL by 30%. Therefore, the causal graph has thoroughly improved the landslide model by incorporating the detected imagery changes from the DPMs that may be classified as a multi-hazard impact of heavy rainfall and ground shaking on soil slopes.

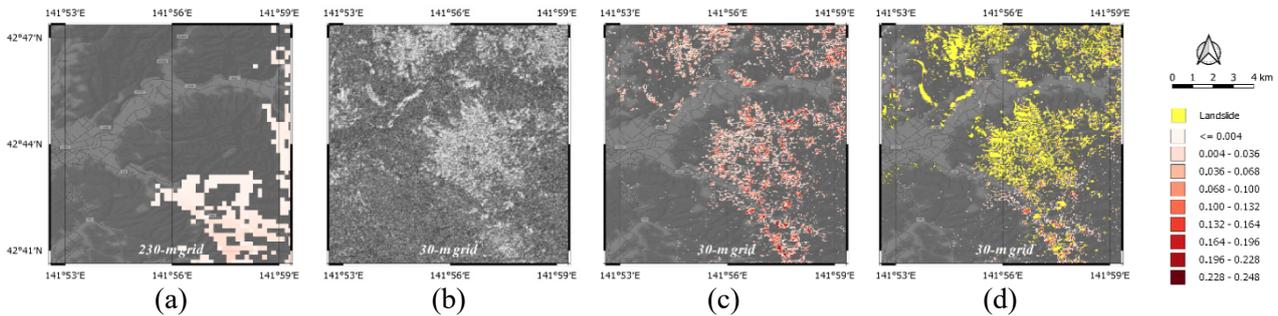

Fig. 2 – (a) Prior and (c) posterior landslide models with (b) DPM and (d) ground truth observations
for the September 2018 Hokkaido Iburi-Tobu earthquake.

In Fig. 3, the prior liquefaction model [10] highlighted areas with significant probability along the river system of Atsuma and Ukuru. However, this prior model with a low 450-m resolution also assigned very small probability to locations with observed landslides (*Fig. 3a*), implying an inaccurate spatial distribution of prior model estimates. To reduce these false positives, the posterior model (*Fig. 3c*) with 30-m resolution has adequately considered the limitations of prior ground failure models in the causal graph through its mutual exclusivity node that allowed the influence of only one ground failure hazard, either landslide or liquefaction. In this case, our analysis did not include the $TPR$, $FPR$, and $CEL$ for liquefaction, because we did not have ground truth data within the given areal extent.

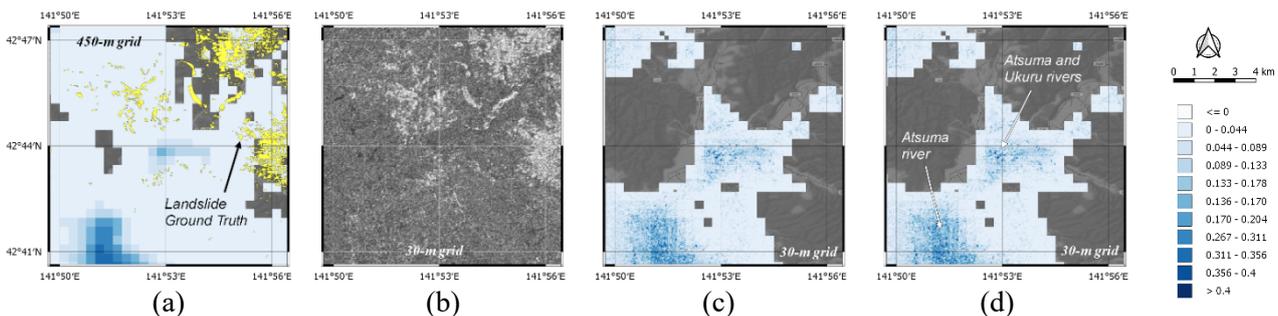

Fig. 3 – (a) Prior and (c) posterior liquefaction models with (b) DPM and (d) locations of water bodies
for the September 2018 Hokkaido Iburi-Tobu earthquake.

Furthermore, the town of Atsuma reported more than 2,960 damaged buildings [31]. In Fig. 4, the posterior model estimated the spatial distribution of these damaged buildings using the building footprints [32] (*Fig. 4a*) and DPM (*Fig. 4b*). This representative posterior model identified priority buildings with relatively high likelihood of severe damage illustrated by the red points (*Fig. 4c*). However, our analysis also did not measure the $TPR$, $FPR$, and $CEL$, because we did not have geotagged data on actual damage to buildings. Note that the USGS continues updating the prior models to provide more accurate estimations. The results shown in Figs. 2, 3, and 4 demonstrate that our model provides accurate posterior distribution of ground failure and building damage even with large uncertainties in the prior models, and as the prior model improves over time, the posterior will improve further.

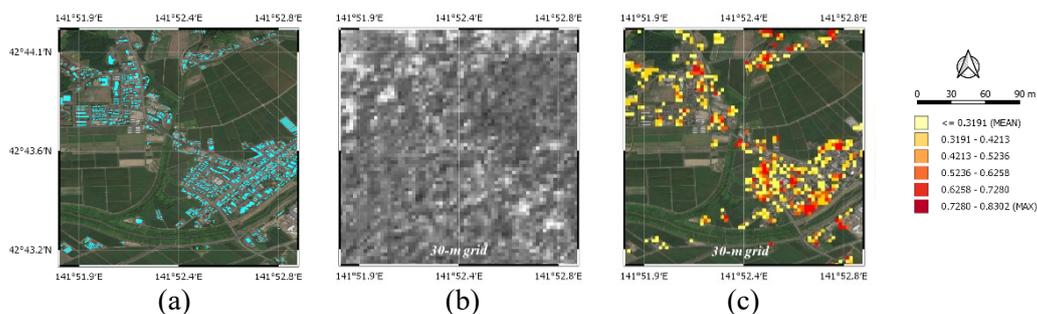

Fig. 4 – (a) Building footprint and (c) posterior building damage model with (b) DPM
for the September 2018 Hokkaido Iburi-Tobu earthquake.

In addition, in Fig. 5, the causal graph has adequately incorporated the probable influence of landslides on the posterior building damage model (*Fig. 5c*). The significant image changes from the DPM (*Fig. 5b*) inferred both posterior landslide and building damage probability in this area (*Fig. 5d*). The reconnaissance efforts also verified these completely damaged buildings were located along steep areas [21].

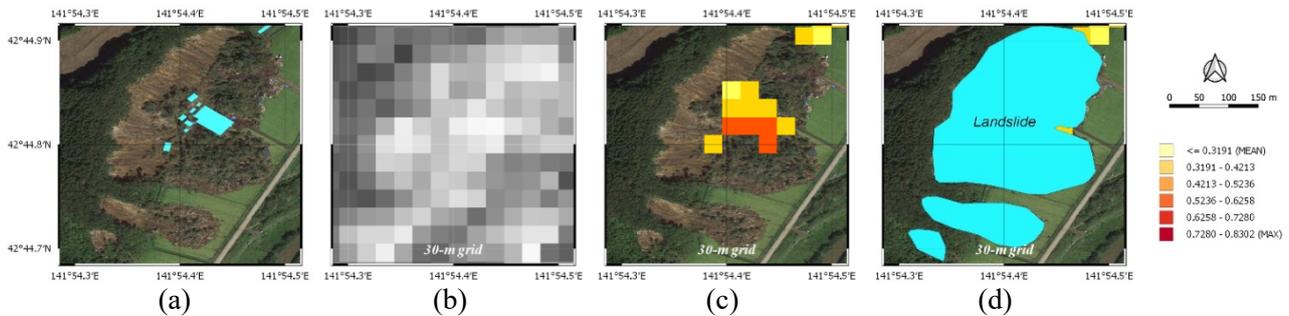

Fig. 5 – (a) An example of identified destroyed house, (b) corresponding DPM with significant ground surface changes detected, (c) posterior building damage estimation, and (d) landslide observations.

## 3.2 January 2020 Southwest Puerto Rico earthquake

A Mw 6.4 earthquake struck the southwest area of Puerto Rico on January 7, 2020, at 4:24 am (AST) [33]. Post-disaster reconnaissance efforts reported widespread casualties with more than 775 affected buildings [34] and 800 ground failure observations [35, 36]. To identify probable damaged areas, the ARIA team generated DPMs using the SAR images from the Copernicus Sentinel-1 satellites of the European Space Agency [37].

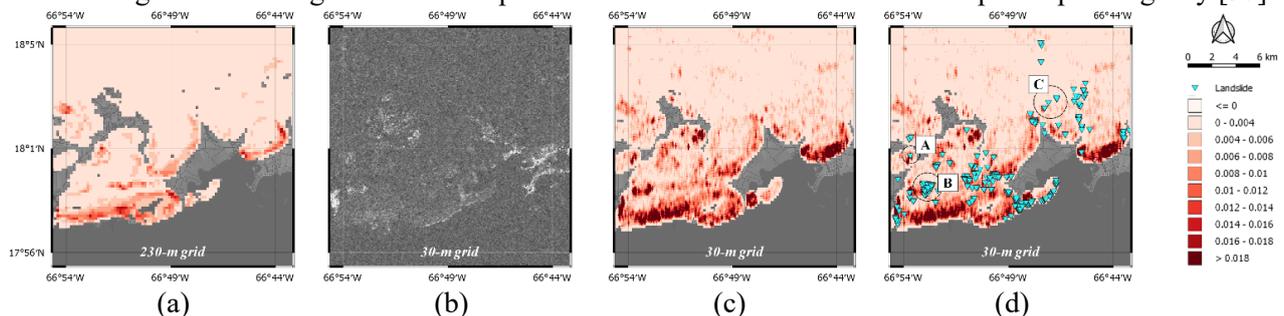

Fig. 6 – (a) Prior and (c) posterior landslide models with (b) damage proxy map and (d) select a

Fig. 6 shows that incorporating the DPM (*Fig. 6b*) reduced the uncertainty in the posterior landslide model (*Fig. 6c*) with a 6% decrease in FPR and a better resolution of at least seven times than that of the prior model (*Fig. 6a*). Fig. 7a also shows that the TPR improved at any thresholds of probability. Ground truth observations such as disrupted slides and falls at 18°00'15"N, 66°53'41"W (A in *Fig. 6d*), boulder at 17°58'54"N, 66°52'44"W (B), and debris at 18°00' 49"N, 66°45'59"W (C) also validated the resulting prominent peaks in the posterior model (*Fig. 6d*). These observations were not observed as peaks in the prior landslide model [26], although it revealed several areas with high probability. Nonetheless, the posterior model reduced the CEL by 52.5%, indicative of improved posterior landslide estimates for the entire area.

Moreover, in Fig. 8, ground cracks with extruded sediments at 17°58'21"N, 66°54'27"W (D in *Fig. 8d*), road fissures due to lateral spreading at 17°58'29"N, 66°48'12"W (E), and flooding due to land subsidence at 18°00'24"N, 66°46'10"W (F) verified the prominent peaks of posterior liquefaction model (*Fig. 8c*) with 15 times higher resolution than that of current prior model [10] (*Fig. 8a*). As shown in Fig. 7b, the posterior model achieves 92.2% TPR with 22.9% FPR, which improves compared to the prior model with 88.5% TPR and 31.8% FPR. The mutual exclusivity node in the causal graph has adequately identified the areas with less probable liquefaction, thereby reducing CEL by 53.3% in posterior model (*Fig. 8d*).

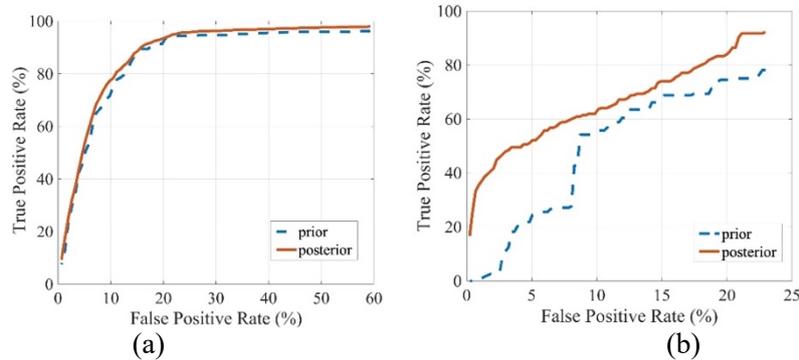

(a)                 (b)

Fig. 7 – ROC curve of prior and posterior models of (a) landslide and (b) liquefaction for the January 2020 Southwest Puerto Rico earthquake.

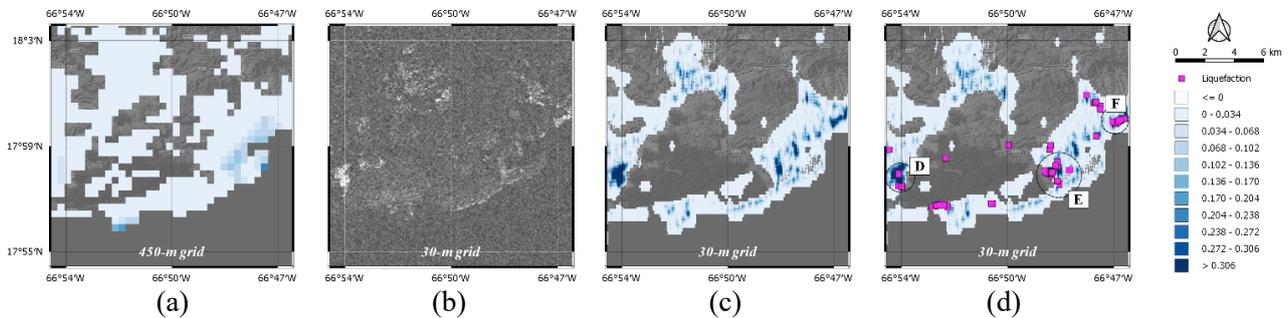

(a)            (b)            (c)            (d)

Fig. 8 – (a) Prior and (c) posterior liquefaction models with (b) damage proxy map and (d) ground truth observations after the January 2020 Southwest Puerto Rico earthquake.

In the town of Guánica, the buildings in the administrative center and the Maria Antonia community suffered substantial damage. Fig. 9 shows that the posterior building damage model predicted the spatial distribution of these damaged buildings, from minor to destroyed, with a TPR of over 76%. In generating this model, the use of the available building footprint data [32] as an input has demonstrated the capability of the causal graph to consider the potential influence of either landslide or liquefaction to building damage.

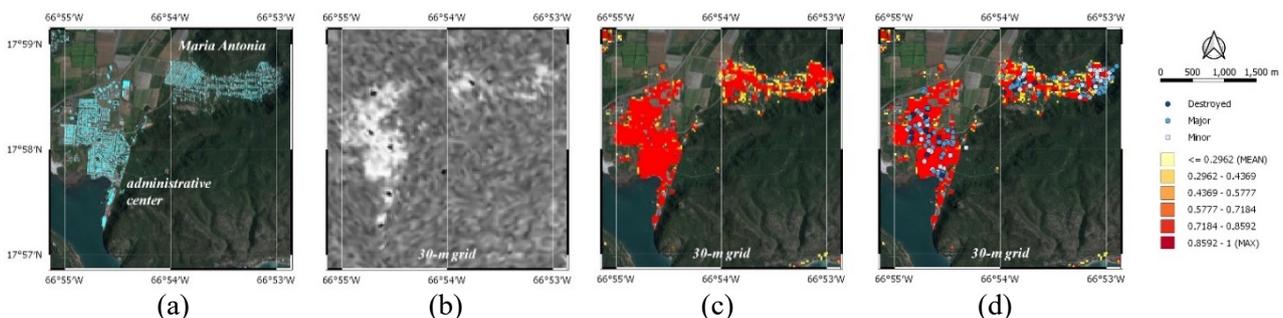

(a)            (b)            (c)            (d)

Fig. 9 – (a) Building footprint and (c) posterior building damage model with (b) damage proxy map and (d) ground truth observations after the January 2020 Southwest Puerto Rico earthquake.

## 4. Conclusions

We present a new joint Bayesian updating framework using a physics-informed causal graph model for post-earthquake ground failure and building damage estimation. The Bayesian causal graph models physical interdependencies among prior models of ground failure, ground failures, building damage, and remote sensing observations. Based on the graph, a stochastic variational inference approach is designed to jointly update the estimations of ground failures and building damage through fusing conventional geospatial models and remote sensing data. We evaluate the algorithm on post-earthquake data collected from Hokkaido, Japan, and Puerto Rico, USA, and compare the estimations with an inventory of observed ground failure and building damage.

The results showed that by incorporating high-resolution imagery, our model significantly reduces the false positive rate of ground failure estimates and improves the spatial accuracy and resolution of ground failure and building damage inferences.

## 5. Acknowledgements

The authors acknowledge the support of NASA ARIA team in providing high-quality updated DPM products. Kate Allstadt and Eric Thompson of the U.S. Geological Survey and Jesse Rozelle of the Federal Emergency Management Agency (FEMA) contributed geospatial ground truth and field observations.